\documentclass[a4paper,10.5pt]{article}

\usepackage[top=1in, bottom=1in, left=1in, right=1in]{geometry}
\usepackage{setspace}
\usepackage{graphicx}
\usepackage{amsmath, amsfonts, amssymb, bbm}
\usepackage{booktabs, longtable, multirow, threeparttable}
\usepackage{siunitx}
\usepackage[inline]{enumitem}
\usepackage{xcolor}
\usepackage{float}
\usepackage{pdflscape}
\usepackage{tikz}
\usepackage{pgfplots}
\pgfplotsset{compat=1.17}
\usepackage{subcaption}
\usepackage{hyperref}
\usepackage{url}
\usepackage{natbib}
\usepackage{amsthm}
\usepackage{appendix}
\usepackage{acronym}
\usepackage{courier}
\usepackage{sectsty}
\usepackage{booktabs}
\usepackage{lscape}

\usetikzlibrary{arrows.meta, positioning, shapes.geometric, fit}
\tikzstyle{provider} = [rectangle, rounded corners, draw=black, fill=yellow!30, text centered, minimum height=1.2cm, minimum width=3.5cm]
\tikzstyle{pool} = [circle, draw=black, fill=purple!30, text centered, minimum size=2cm]
\tikzstyle{arrow} = [thick, ->, >=stealth]
\tikzstyle{decision} = [diamond, draw=black, fill=green!30, text centered, minimum height=1.5cm, minimum width=3cm]
\tikzstyle{fee} = [trapezium, trapezium left angle=120, trapezium right angle=60, draw=black, fill=teal!30, text centered, minimum width=2cm, minimum height=1cm]
\tikzstyle{action} = [rectangle, draw=black, fill=gray!20, text centered, minimum height=1.2cm, minimum width=4cm]

\AtBeginEnvironment{tabular}{\scriptsize}
\AtBeginEnvironment{longtable}{\scriptsize}
\AtBeginEnvironment{longtable}{\singlespacing}
\usepackage[labelfont=bf,font={scriptsize},justification=justified,singlelinecheck=false]{caption}

\sectionfont{\normalsize\bfseries}
\subsectionfont{\normalsize\itshape}
\subsubsectionfont{\normalsize\itshape}
\paragraphfont{\normalsize\itshape}

\hypersetup{
    hidelinks,
    colorlinks=true,
    citecolor=blue,
    linkcolor=blue,
    urlcolor=blue,
    pdfmenubar=false,
    pdftoolbar=false
}

\newcommand{\email}[1]{\href{mailto:#1}{\texttt{#1}}}

\begin{document}

\title{\Large \textbf{Institutionalizing risk curation in decentralized credit}}
\author{
  \scriptsize Anastasiia Zbandut\\ \email{anastasiia@blockworks.co}
  \and 
  Carolina Goldstein\\ \email{carolina@blockworks.co} 
}
\date{\small This draft: December 12, 2025}
\maketitle
\begin{abstract}
This paper maps the emerging market for decentralized credit in which ERC--4626 vaults and third-party curators, rather than monolithic lending protocols alone, increasingly determine underwriting and leverage decisions. We show that modular vaults differ in capital utilization, cross-chain and cross-asset concentration, and liquidity risk structure. 
Further, we show that a small set of curators intermediates a disproportionate share of system TVL, exhibits clustered tail co-movement, and captures markedly different fee margins despite broadly similar collateral composition. These findings indicate that the main locus of risk in DeFi lending has migrated upward from base protocols, where underwriting is effectively centralized in a single DAO-governed parameter set, to a permissionless curator layer in which competing vault managers decide which assets and loans are originated.
We argue that this shift requires a corresponding upgrade in transparency standards and outline a simple set of onchain disclosures that would allow users and DAOs to evaluate curator strategies on a comparable, money-market–style basis.
\end{abstract}


\section{Introduction}

Between 2022 and 2025, DeFi underwent a structural transformation comparable in scope to the development of money-market funds in the 1970s. Protocol-governed credit systems such as Aave and Compound, which previously centralized both liquidity management and risk calibration, have evolved into modular architectures where vaults take on distinct risk profiles and third-party curators determine underwriting parameters, leverage bounds, and oracle feeds.  

This transition, enabled by the ERC--4626 vault standard and formalized through platforms such as Morpho, Euler, Silo, and Gearbox Permissionless, separates infrastructure from risk management. The empirical question is no longer which protocol dominates TVL, but how efficiently modular lending transforms liquidity into yield, and how curator specialization reshapes systemic risk propagation.

Empirically, existing DeFi lending does not resemble bank-style credit screening but rather a standardized, overcollateralised leverage machine. \citet{lockedin2025risk} formalize this view by modeling borrowers as dynamically rebalancing leveraged positions that are "locked in" by protocol collateral rules and liquidation mechanics. In their framework, borrowers optimally lever up to exploit convex payoff profiles, but face a sharply non-linear ruin probability once volatility or funding conditions deteriorate, so that realized returns are highly skewed and concentrated in a small set of surviving paths. Using Aave V2 data, \citet{cornelli2025defilending} show that depositors are predominantly retail investors engaging in search-for-yield behavior, while borrowing is driven by speculative and governance motives, with WETH collateral backing stablecoin liabilities in most cases. 
Our analysis characterizes the resulting two-layer architecture, in which base lending protocols provide shared market infrastructure and accounting, while curator-managed ERC--4626 vaults supply distinct risk profiles and active risk management, and studies how credit intermediation and systemic risk are distributed across protocols, chains, and curators.

We analyze onchain data with the time horizon from October 1, 2024 until November 19, 2025 for six major lending systems, i.e., Aave (V2 and V3), Morpho (Blue), Euler (V2), Maple, Gearbox, and Silo (V2), and for eight large curators, i.e., Gauntlet, Steakhouse, MEV Capital, K3 Capital, R7, Block Analitica, Yearn, and B Protocol.  
We characterize the lending layer through three dimensions:  
(i) utilization and yield scalability,  
(ii) concentration and chain diversification, and  
(iii) curator network topology and contagion potential.  
Finally, we link these findings to analogs in regulated finance, highlighting the convergence between DeFi curators and traditional collateral managers, UCITS risk units, and CLO structures.

This paper makes three contributions.
First, we document the transition from monolithic lending protocols to a two-layer credit architecture in which ERC--4626 vaults and third-party curators play a dominant role. Second, we characterize the curator layer as a network of heterogeneous but interconnected balance sheets. We show how portfolio composition, factor concentration, and co-movement of liquidity stress jointly determine which curators act as systemic hubs. Third, we use these measurements to motivate a minimal transparency standard for modular credit. We translate familiar prudential concepts into onchain disclosure items that can be implemented using existing ERC--4626 primitives and indexing infrastructure.

The remainder of the paper proceeds as follows. Section~\ref{sec:delegated-risk} describes the shift from standardized pools to delegated risk and positions modular vaults within the broader DeFi lending landscape. Section~\ref{sec:efficiency-yield} analyzes utilization and yield dynamics across vault architectures. Section~\ref{sec:systemic-concentration} studies execution-layer concentration and TVL co-movement in vaults. Section~\ref{sec:curator-networks} turns to the curator layer, documenting portfolio structure, tail dependence, and network centrality. Section~\ref{sec:transparency} proposes an onchain transparency framework, and Section~\ref{sec:conclusion} concludes the paper.

\section{From standardized vaults to delegated risk}
\label{sec:delegated-risk}

The ERC--4626 standard (2023) created a uniform accounting layer for yield-bearing vaults, allowing deposits and redemptions to function as fungible shares. Within months, adoption by Yearn, Balancer, Frax, and Sommelier triggered a vault proliferation analogous to the ETF boom: capital flowed into modular wrappers rather than protocol pools.  

Morpho (2024) institutionalized this modularity by introducing risk-managed vaults that function as non-custodial "model portfolios" of overcollateralised lending positions. On the vault layer, curators specify a risk profile and target allocations across underlying lending markets that depositors can enter and exit at any time, while the base protocol continues to provide the generic market and accounting infrastructure. Where suitable markets already exist, vaults allocate into them directly and, where they do not, curators may deploy new markets on the base layer, effectively underwriting additional credit commitments within the same standardized infrastructure.
Euler’s vault kit, Silo’s isolated markets, and Gearbox’s permissionless lending achieved the same outcome where risk became local, composable, and curator-defined. 

In the classical DeFi lending protocols, risk is encoded in a small set of global parameters (loan-to-value caps, liquidation thresholds and interest-rate curves) that are updated only infrequently through token-holder governance. \citet{ChiuEtAl2024Fragility} document that Aave’s lending activity is highly volatile and highlight concerns about procyclicality with respect to crypto prices. They also show that key risk parameters were changed only 13 times in the first two years of Aave V2, despite large swings in volatility and collateral composition, illustrating how rigid protocol-level risk rules can be.
For a fixed configuration of fundamentals, the protocol can coordinate either on a high-lending "pooling" equilibrium or on a low-lending "separating" equilibrium with depressed collateral prices and tight effective haircuts, so that "market runs" emerge from expectations about future collateral valuations rather than from traditional deposit withdrawals. Delegating risk management to an entity that can adjust haircuts flexibly in response to market conditions restores a unique, high-lending equilibrium in their model, but at the cost of introducing a centralized risk manager.

\citet{HeimbachEtAl2024ShortSqueeze} analyze the November 2022 CRV short-squeeze attempt on Aave V2. They reconstruct Avi Eisenberg’s on- and offchain trades and show that, because roughly one-third of CRV’s circulating supply was available to borrow on Aave, a single failed attack was sufficient to generate more than USD~1.5~million in bad debt and to trigger the freezing of several long-tail assets. Their discussion highlights a structural trade-off for large monolithic lending protocols. Either restrict the universe of supported assets and cap each token’s borrowable share of market capitalization, or appoint an active "risk admin" who can rapidly adjust risk parameters at the expense of weaker decentralization. 

While Aave extended its cross-chain reach, it retained DAO-governed, uniform risk parameters, thereby prioritizing solvency over marginal capital utilization, but also giving up risk specific customization. 
In a modular architecture, different user segments can express heterogeneous risk preferences through curator-defined vaults, whereas Aave’s one-size-fits-all grid leaves collateral in shared pools without hard isolation and with largely uniform pricing of collateral risk. This prevents fine-grained matching between borrower types and risk profiles and implicitly induces cross-subsidization across users.
In theory, such segmentation is one of the most valuable features of vault-based systems, but today users lack clear visibility into the specific risk profile of each vault. The absence of this transparency undermines the benefit and can even turn customization into a liability. 

The result is a structural dichotomy. Monolithic protocols deliver liquidity depth and governance coherence, while modular vaults deliver specialization, faster responsiveness, and higher efficiency per unit of TVL.  
The remaining question, empirical rather than ideological, is which design allocates credit more efficiently and distributes risk more transparently.

Existing empirical work on DeFi lending has so far focused mainly on monolithic protocols. For Compound~V2, \citet{tovanich2023contagion} reconstruct pool-level balance sheets from onchain data and model the protocol as a network of interlinked liquidity pools. They show that most users either borrow stablecoins against crypto collateral or engage in liquidity mining, and that contagion is driven by default cascades running from crypto-collateral pools to stablecoin pools. In their stress tests, stablecoin pools are the most likely to default, while crypto pools are the primary sources of systemic shocks.
The same evidence suggests that these lending pools do not channel capital into heterogeneous credit exposures but mostly into leveraged crypto risk. \citet{cornelli2025defilending} show that more than 98\% of Aave users maintain a health factor above one and that borrowing is primarily driven by speculative and, for large users, governance-related motives. Borrowing volumes rise significantly with ETH perpetual futures funding rates and past ETH returns, while borrowing of governance tokens spikes around protocol voting dates. In other words, standardized DeFi money markets mainly connect yield-seeking depositors with leveraged speculators in liquid onchain assets, rather than with idiosyncratic project risk.

Recent research has started to examine decentralized credit markets at the protocol level, combining credit-risk and governance perspectives. \citet{Oyeyemi2025DecentralizedCredit} study MakerDAO, Aave, Compound, Maple, and Goldfinch and construct a multi-factor framework that links collateral characteristics, liquidation frequencies, and a DAO governance effectiveness index to realized liquidation events. Their logistic regressions suggest that higher loan-to-value ratios and collateral volatility increase liquidation risk, while stronger governance scores are associated with fewer liquidations and better crisis management. Complementing this protocol-level perspective, \citet{lockedin2025risk} derive a borrower-level risk–return–ruin frontier for overcollateralised DeFi loans, showing how leverage choices interact with protocol collateral rules to generate fat-tailed loss distributions. 
\citet{PackinLevAretz2024BlackBox} analyze decentralized credit scoring as a "black box 3.0" problem, highlighting how opaque, algorithmically mediated scores can import bias, fairness concerns, and consumer-protection risks into DeFi credit allocation wherever access to lending rails is gated by on- or offchain scoring mechanisms.
Our contribution is to move one layer up, from protocol balance sheets, borrower-level leverage choices, and credit-scoring design to the modular vault and curator layer that increasingly intermediates DeFi credit across multiple lending venues and chains.

\section{Utilization and yield dynamics}
\label{sec:efficiency-yield}

\begin{figure}[!htbp]
\centering
\includegraphics[width=\textwidth]{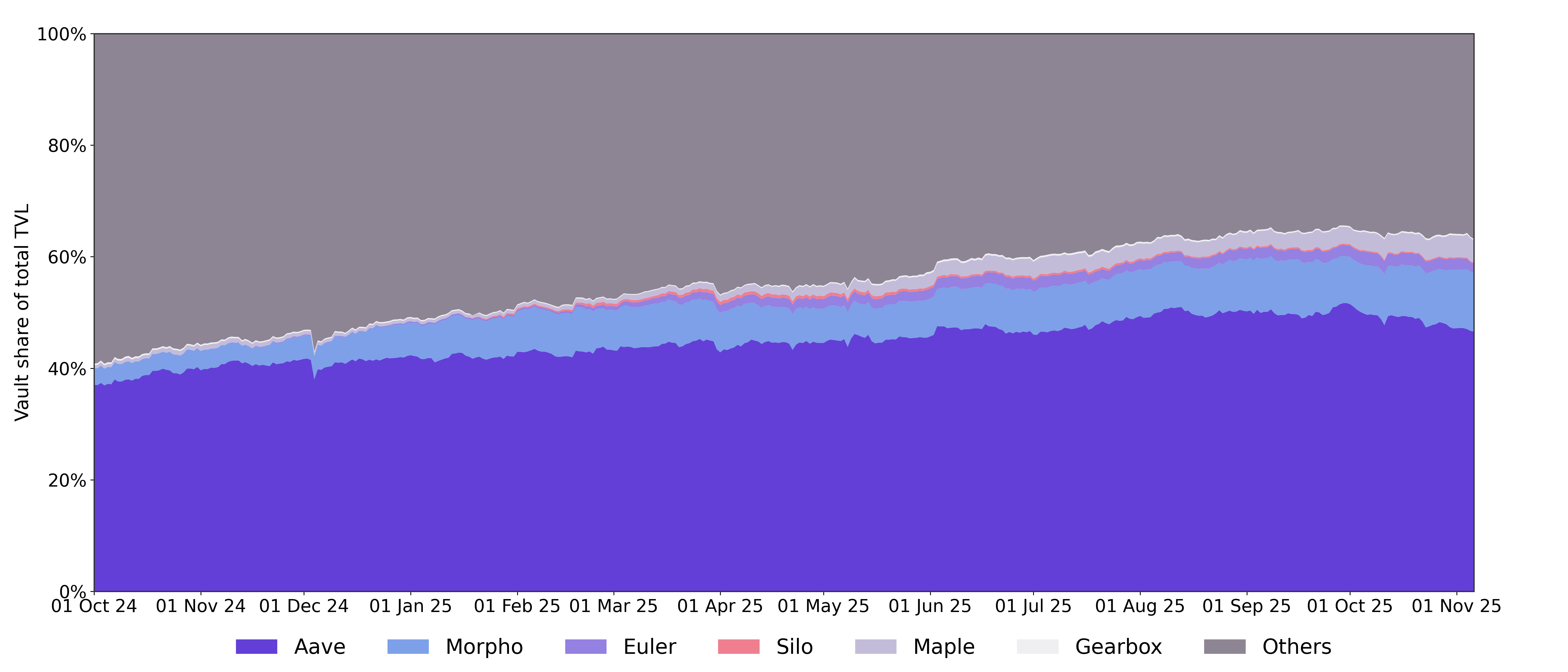}
\caption{Market share of total lending TVL among major protocols.}
\label{fig:share_tvl_lending}
\end{figure}

Figure \ref{fig:share_tvl_lending} traces aggregate TVL shares across the leading protocols. Aave remains the largest universal lending protocol, jointly holding approximately \$34~billion in TVL, or 45\% of total sector liquidity. Yet, relative to the growth of aggregate lending TVL, Aave’s dominance is essentially unchanged. Its market share remains broadly stable, implying that the gains of modular vaults have come largely at the expense of the residual “others” segment. This implies that competitive dynamics have played out primarily among smaller protocols, with Aave retaining its role as the system’s primary liquidity anchor. 

Capital utilization (defined as the ratio of active loans to TVL) differentiates lending architectures sharply. In the terminology of \citet{lockedin2025risk}, higher utilization pushes systems closer to the "locked-in" region of the risk–return–ruin frontier, where incremental fee income is traded off against a disproportionately higher probability of borrower liquidation and protocol losses.
In practice, this mapping from utilization to risk is mediated by the choice of interest-rate model. Most DeFi money markets still employ some variant of the original two-sloped “kinked” curve with an optimal-utilization point (typically around 90\%), beyond which the marginal cost of borrowing rises steeply. As a result, small shifts in utilization near the kink can induce disproportionately large changes in funding costs and in the stability of high-utilization regimes.

\begin{figure}[!htbp]
\centering
\includegraphics[width=\textwidth]{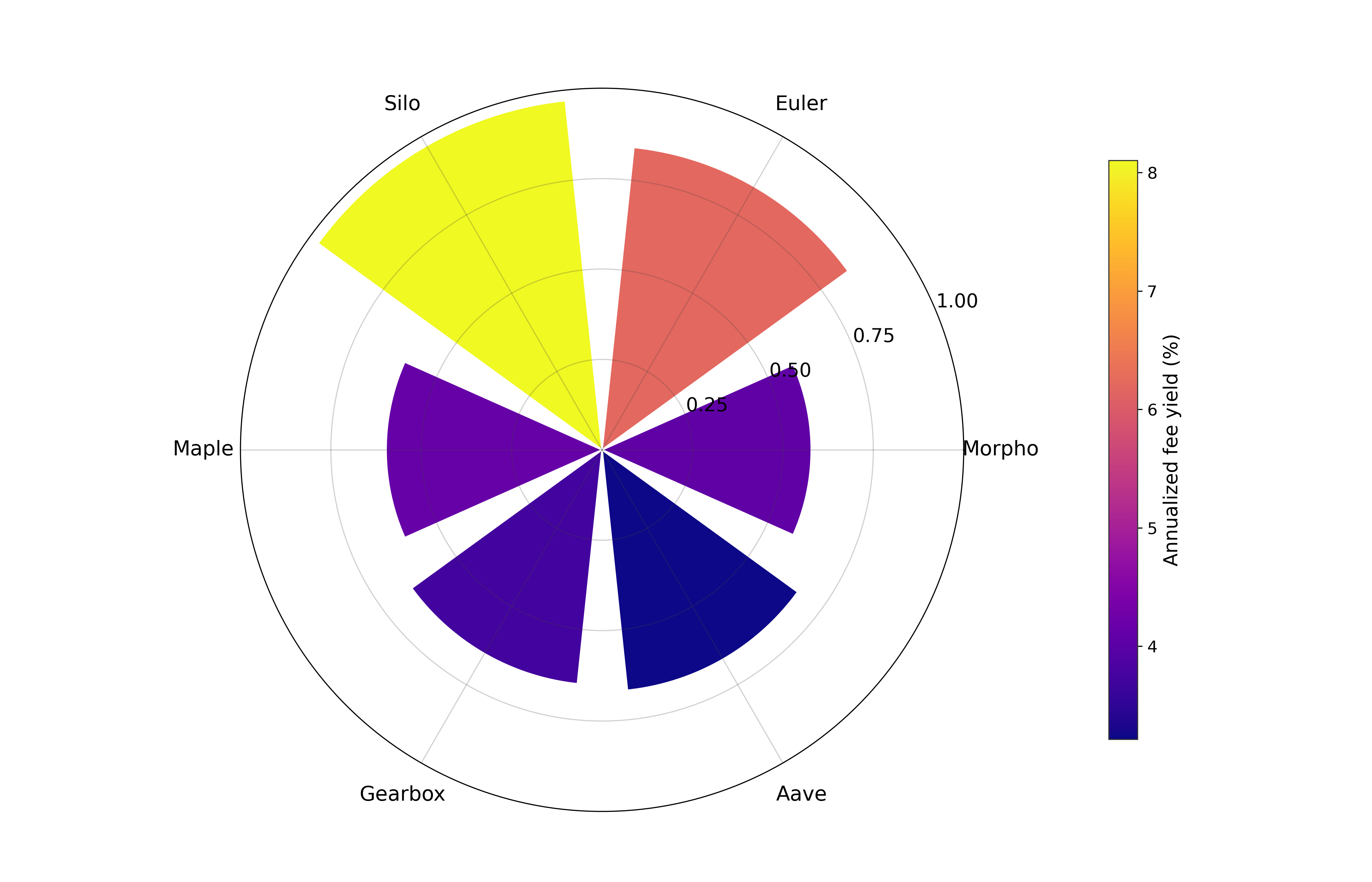}
\caption{Risk–yield frontier.}
\label{fig:risk_yield_frontier_vaults}
\end{figure}

Figure~\ref{fig:risk_yield_frontier_vaults} positions each protocol along the empirical utilization–yield frontier, where the radius represents mean capital utilization and color encodes annualized mean fee yield. The relationship is monotonically increasing across the sample. Protocols that keep a larger share of their balance sheet deployed also realize higher fee income. Our frontier is measured at the protocol balance-sheet level (fees versus utilization), whereas \citet{lockedin2025risk} derive a borrower-level risk–return–ruin frontier for overcollateralised loans. While the two frontiers are not directly comparable, they are conceptually complementary. Empirically, we show that higher utilization is associated with higher fee yields at the protocol level, and \citet{lockedin2025risk} show that, for a given lending design, higher leverage at the user level is associated with a more skewed, ruin-prone loss distribution.

Silo and Euler sit closest to the empirical frontier, with mean capital utilization of roughly 0.97 and 0.84 and annualized fee yields of about 8.1\% and 6.2\%, respectively. Their near-full utilization is enabled by isolated-market architecture that confines liquidation risk to each vault’s specific collateral set. Therefore, curators can push loan-to-value and interest-rate parameters further without exposing the rest of the system to spillovers.

Maple and Gearbox cluster in the intermediate region (capital utilization around 0.60-0.65, fee yields 3.7-4.1\%). In Maple, loans are centrally underwritten and structured as overcollateralised, offchain credit exposures to institutional borrowers, so utilization can be high, but yields are tempered by internal credit screens and loss-buffer mechanisms.
Gearbox operates overcollateralised credit accounts that rehypothecate borrowed assets into external DeFi venues. Gearbox's risk engine enforces conservative margin and liquidation rules across a diversified set of whitelisted strategies, which keeps it away from the extreme corner of the frontier.

Aave and Morpho occupy a more conservative band (capital utilizaion about 0.58 – 0.67, fee yields 3.2 – 4.1\%), consistent with designs that prioritize solvency and broad user safety over maximum turnover. Aave relies on DAO-governed, protocol-wide risk parameters to set collateral factors and liquidation thresholds, which delivers a robust "one-size-fits-most" profile but limits risk-segment differentiation. Morpho's vault layer allows curator-specific parameterization, but collateral posted to Morpho vaults is not rehypothecated by default and remains available on the smart contracts for liquidation, and the protocol is explicitly designed such that it cannot become balance-sheet insolvent. In practice, the largest vaults are ETH–stablecoin markets managed by relatively conservative curators such as Gauntlet and Steakhouse, so the aggregate footprint remains close to Aave's.

\section{Systemic concentration and interdependence}
\label{sec:systemic-concentration}

\begin{figure}[!htbp]
\centering
\includegraphics[width=\textwidth]{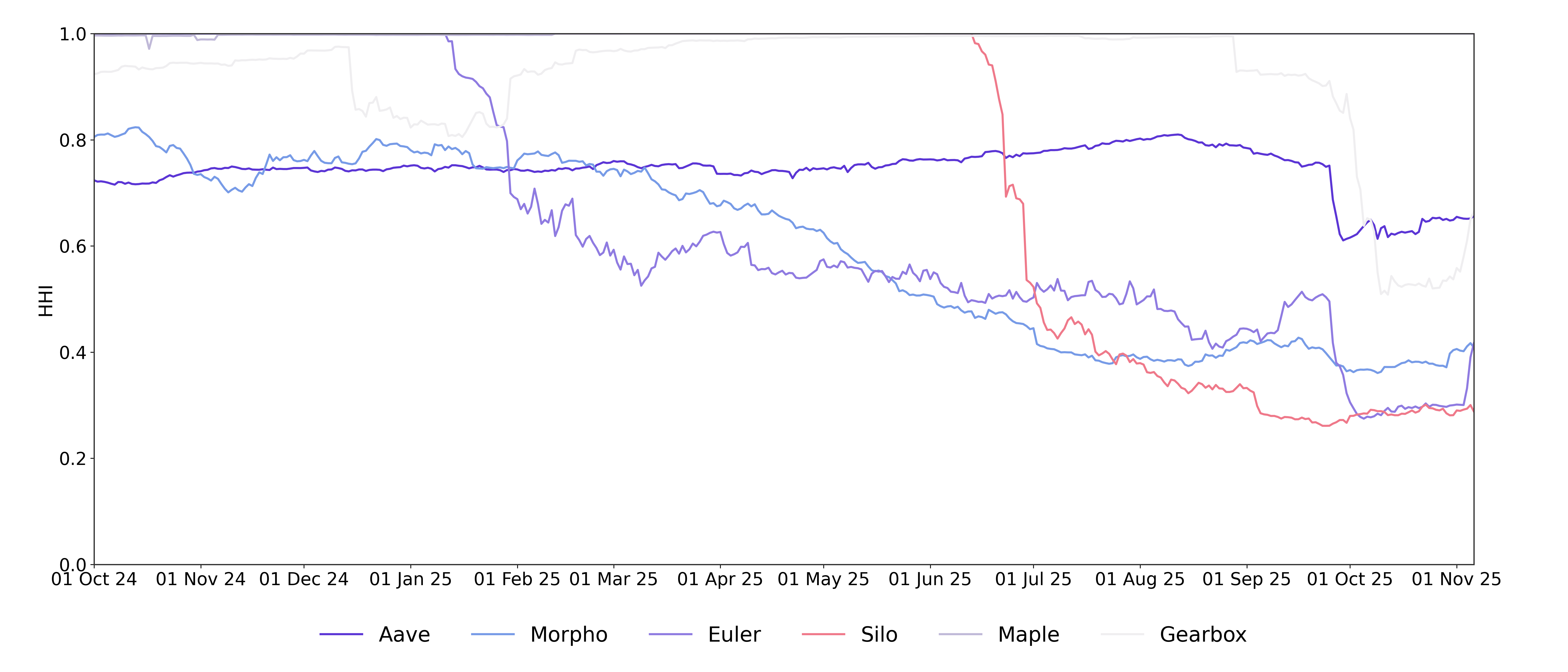}
\caption{Evolution of chain concentration.}
\label{fig:chain_concentration_vault}
\end{figure}

Systemic resilience depends not only on utilization, but also on how lending liquidity is distributed across execution layers. Our approach is complementary to protocol-specific contagion studies. \citet{tovanich2023contagion} use pool-level balance sheets to simulate default cascades and show that crypto-collateral pools can transmit shocks to stablecoin pools through interpool liabilities. In contrast, we study co-movement and concentration across multiple lending architectures and chains, treating protocol TVL and chain Herfindahl–Hirschman Indices (HHIs) as aggregate state variables.
The HHI in Figure~\ref{fig:chain_concentration_vault} is computed from each protocol’s TVL shares by chain and summarizes execution-layer concentration. Values close to 1 indicate full concentration on a single execution layer, whereas lower values reflect dispersion across multiple chains. As such, the HHI provides a compact measure of execution-layer diversification and highlights differences in operational risk exposure across lending architectures. 

Aave exhibits the most stable execution-layer concentration profile in the sample. Despite operating on the broadest multichain footprints, i.e., Ethereum, Avalanche, Polygon, Base, Fantom, OP Mainnet, Scroll, Metis, Gnosis, BSC, ZKsync Era, and Arbitrum, followed by sequential deployments to Linea (mid-Feb 2025), Sonic (early Mar 2025), Celo (mid-Apr 2025), Sone-ium (early Jun 2025), and Plasma (late Sep 2025), its TVL distribution remains persistently top-heavy. The HHI floats around 0.70 for most of the observation window, indicating that Ethereum and a small cluster of L2s continue to dominate liquidity despite additional chain integrations. A sharp decline to $\approx$ 0.60 occurs in October 2025, reflecting the first dispersion of assets across the newly added execution environments. After this one-time adjustment, Aave’s HHI stabilizes around 0.60, suggesting that marginal TVL on new chains is diversifying the footprint without meaningfully altering the protocol’s structural anchoring on Ethereum and a few major L2s.

Morpho declines smoothly from $\approx$ 0.80 in late 2024 to ~0.35 – 0.40 by October–November 2025, a pattern fully explained by its staged multichain rollout. It started as Ethereum-only, adding Base in mid-June 2024, then Fraxtal in February 2025, followed by Polygon, Arbitrum, OP Mainnet (mid-March), Sonic (early April), World Chain and Corn (late April), Hyperliquid (early May), Plume (late May), Unichain (early June), Lisk (late June), Katana (early July), TAC (mid-July), Soneium (late August), Hemi (late September), and finally Sei, Botanix, Scroll, and Etherlink in October 2025. This sequential onboarding pushes the HHI lower as TVL disperses across an expanding set of execution layers.
Euler follows a similar downward path, falling from $\approx$ 0.90 at the start of 2025 to the 0.30 – 0.40 range by late summer, mirroring its sequential multichain rollout. Initially Euler was Ethereum-only, expanding to Base and SwellChain (January), Sonic and BOB (late February), Berachain (mid-March), Avalanche and BSC (mid-April), Unichain (mid-May), Arbitrum (late June), TAC (mid-July), Linea (mid-August), and Plasma (late September). Each onboarding event introduces a new TVL venue, dispersing liquidity and reducing the protocol’s chain concentration.

Silo's concentration drops to its period low just above 0.40 and then stabilizes near that level. The sharp fall in the late June-early July is due to TVL migration from Sonic-only to genuinely multichain deployment. As Silo's lending markets on Arbitrum, Ethereum and Avalanche begin to accumulate TVL balances, Sonic's share of total TVL declines and chain weights become more evenly distributed. This produces the steepest one-time deconcentration in the sample, reflecting Silo's rapid expansion phase.
Gearbox remains highly concentrated for most of the sample window because activity is initially limited to Ethereum (major), Arbitrum, and OP Mainnet. The concentration ratio only begins to break in late summer as the protocol expands to Lisk, Sonic, Hemi, Etherlink, and for a short time to BSC in late August 2025, followed by Plasma (late September). This rapid burst of multichain onboarding drives the sharp HHI drop toward $\approx$ 0.50, while the partial rebound to $\approx$ 0.60 in the following month reflects early-stage and uneven liquidity distribution across the newly added chains.

Maple remains close to an HHI of 1.0 across the sample, consistent with its predominantly single-chain footprint. Although it was deployed on Solana and Ethereum, meaningful liquidity and lending activity is based on Ethereum chain.
Overall, these paths show heterogeneous trade-offs between operational simplicity (high HHI) and resilience to chain-specific shocks (low HHI).

\begin{figure}[!htbp]
\centering

\begin{minipage}{0.48\textwidth}
  \centering
  \includegraphics[width=\textwidth]{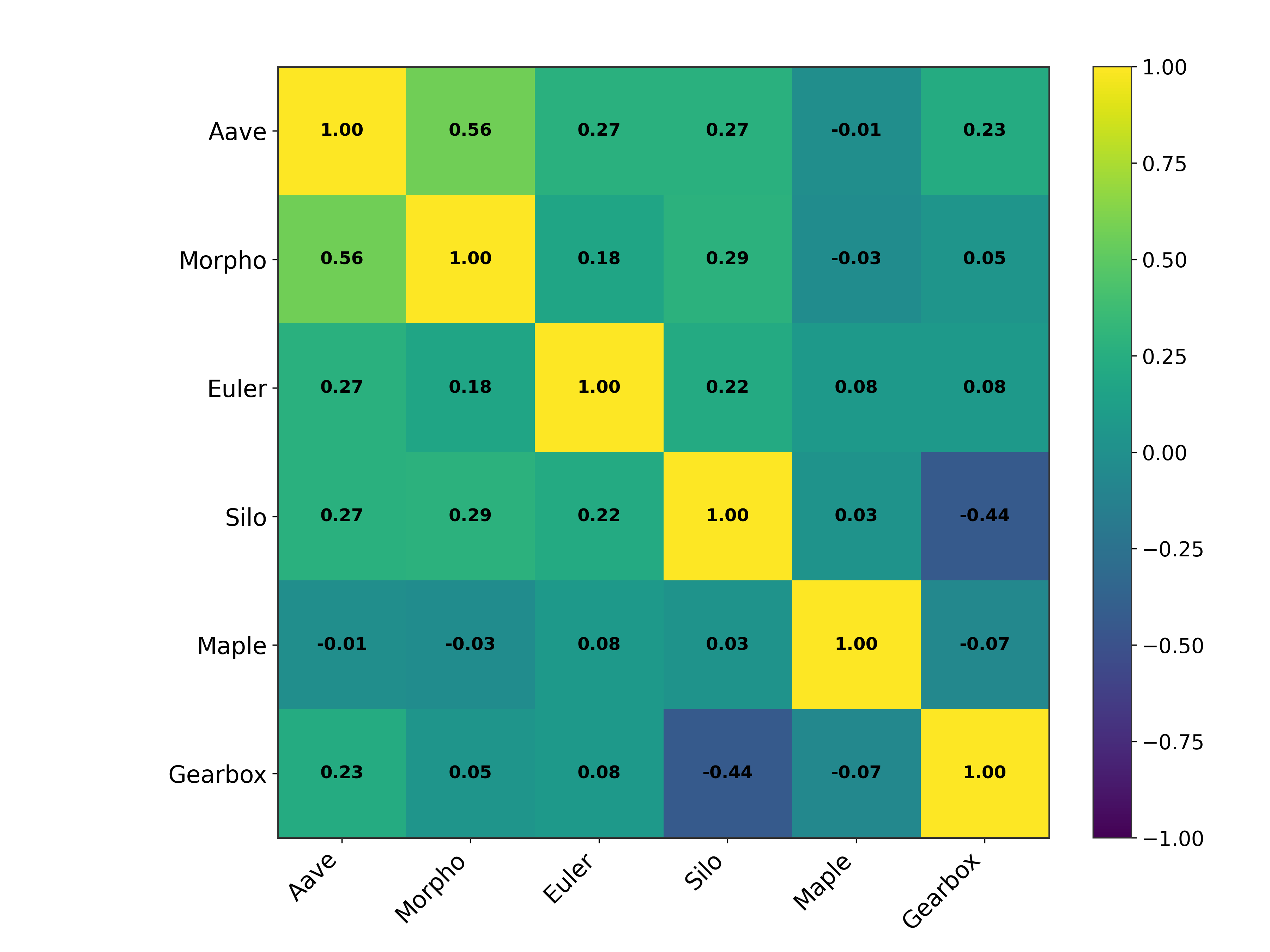}
  \vspace{0.3em}
  {\tiny Sample 1:\ 01.10.2024--19.04.2025}
\end{minipage}
\hfill
\begin{minipage}{0.48\textwidth}
  \centering
  \includegraphics[width=\textwidth]{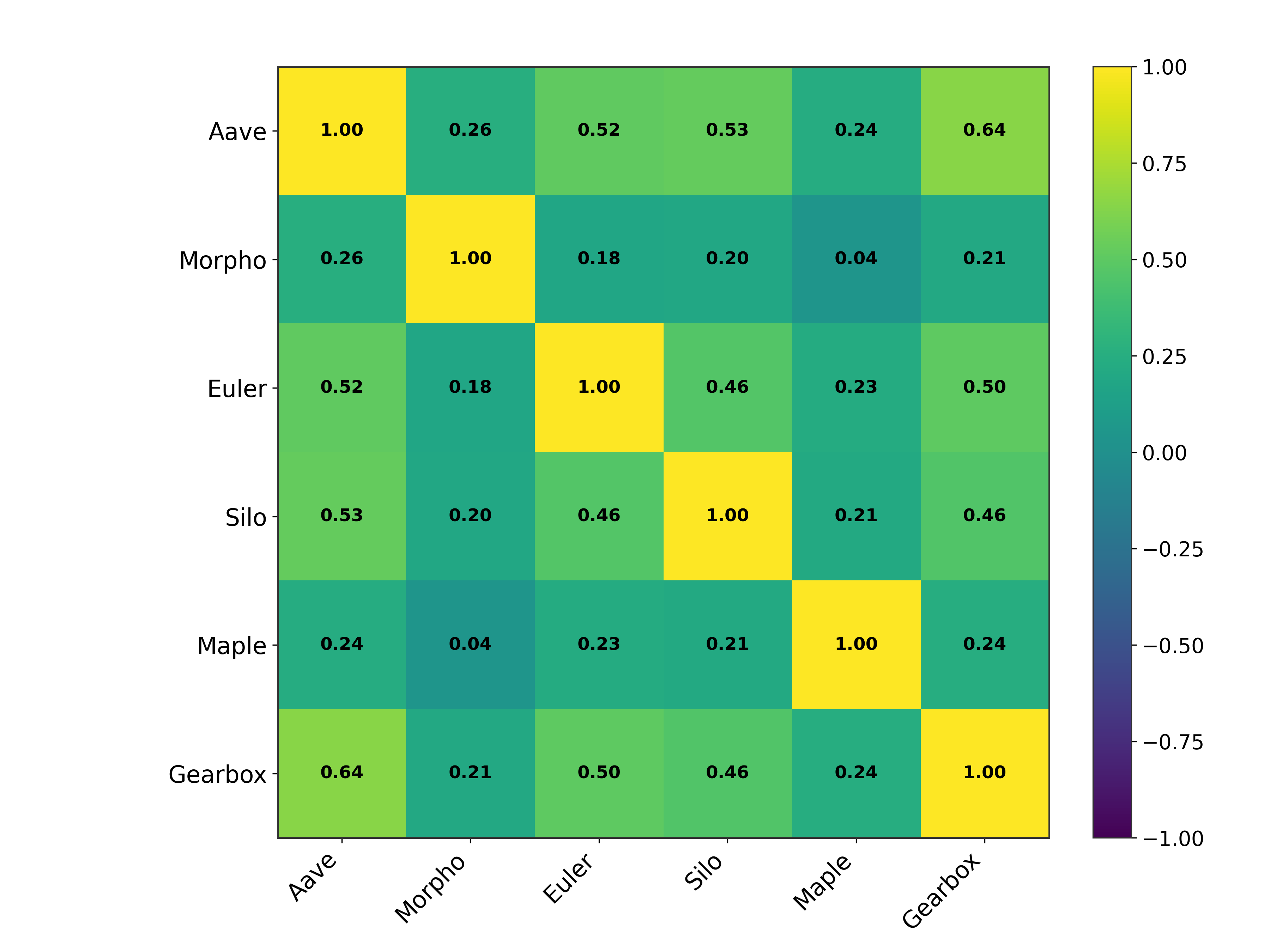}
  \vspace{0.3em}
  {\tiny Sample 2:\ 20.04.2025--06.11.2025}
\end{minipage}

\caption{Correlation matrices of daily TVL changes.}
\label{fig:correlation_tvl_vault}
\end{figure}

To show how relationships among protocol TVLs evolve over time, Figure~\ref{fig:correlation_tvl_vault} reports correlation matrices of daily TVL changes for two equal-length subsamples. In Sample~1 (01.10.2024 – 19.04.2025), correlations are low and heterogeneous. Most coefficients lie between -0.45 and +0.30. Only Aave exhibits moderate linkages with Morpho (0.56), Euler and Silo (both $\approx$ 0.27), whereas Maple remains effectively uncorrelated with the rest of the sector. The Gearbox–Silo pair stands out with a strong negative coefficient (-0.44), indicating capital rotation between leveraged execution environments and isolated collateral pools rather than synchronized liquidity cycles.

In Sample~2 (20.04.2025 – 06.11.2025), the entire correlation structure shifts upward and becomes uniformly positive. Aave’s co-movement with Euler, Silo, and Gearbox increases sharply to $\approx$ 0.52, 0.53, and 0.64, respectively. Maple’s correlations rise across the board but remain weak in absolute magnitude ($\approx$ 0.20 – 0.25), consistent with its narrower institutional borrower base. By contrast, Morpho’s correlations slightly decline relative to the first subsample, e.g., Aave–Morpho from 0.56 to 0.26 and Morpho–Silo from 0.29 to 0.20, reflecting the persistence of idiosyncratic vault-level credit cycles despite sector-wide convergence.

Three structural mechanisms account for the pronounced increase in overall correlation. 
First, the major multichain deployments of Morpho, Euler, and Gearbox are largely completed by mid-2025, reducing measurement noise from TVL fragmentation across newly activated execution layers. As redeployments stabilize, daily TVL variation increasingly reflects aggregate liquidity conditions rather than protocol-specific expansion cycles.  
Second, collateral composition homogenizes over time as ETH- and stablecoin-backed markets become the dominant drivers of TVL across all architectures. Greater similarity in underlying asset exposure amplifies the responsiveness of flows to macro-level shifts in ETH-beta and stablecoin demand.  
Third, the broad liquidity wave associated with the 2025 L2 expansion (Base, Lisk, Sonic, Arbitrum, and others) induces synchronized inflows across lending markets integrated into these ecosystems, creating a common-cycle component in TVL dynamics.

The resulting matrix in the second subsample reflects a transition from an architecture-segmented regime toward a more unified liquidity environment. Aave, Euler, Silo, and Gearbox begin to share a strongly positive common factor, Maple remains only weakly connected, and Morpho (despite modestly lower correlations) continues to exhibit structurally distinct modular-vault dynamics. Together, the two subsamples illustrate the sector’s evolution from idiosyncratic protocol-specific credit cycles toward a more synchronized, multichain liquidity regime shaped by collateral convergence and coordinated L2-driven capital flows.

\section{Curator networks and contagion channels}
\label{sec:curator-networks}

Delegation of risk to third-party managers introduces a second systemic layer in the form of curator interdependence. Methodologically, our construction parallels \citet{tovanich2023contagion} work that treats lending pools as nodes in a financial network with interpool claims. Whereas their study analyzes contagion within Compound by simulating liquidations and pool defaults, we build an overlap network at the curator layer, where nodes represent managers and edges capture shared exposures across protocols, chains, and vaults.

Our unit of analysis is the curator as an aggregate balance sheet, not the individual vault. This aggregation inevitably suppresses some within-curator heterogeneity, where a single curator can operate multiple vault product lines with distinct mandates and risk profiles, e.g., low-volatility “cash-plus” vaults alongside higher-beta strategies. In such cases, overlapping curator footprints need not imply that any given pair of vaults is closely connected in risk space. Accordingly, we interpret the network as capturing shared systemic footprint at the curator level, rather than substitutability or direct contagion between specific vaults, and leave a fully vault-level overlap network to future work.

Although vaults are isolated by design, this isolation applies primarily to credit rather than liquidity risk. In practice, curators frequently hold overlapping asset sets and route positions through the same underlying lending markets, so that capital from multiple vaults is aggregated at the execution layer. This aggregation generally improves depth and pricing relative to fully siloed, per-vault markets, but it also means that liquidity conditions are shared. When one vault faces redemptions due to idiosyncratic credit events, it may be forced to unwind otherwise unaffected positions, temporarily draining liquidity and pushing up interest rates in shared markets. In such episodes, lenders in other vaults that only hold the unaffected side of these markets can earn outsized risk–return premia: credit risk remains localized, but liquidity stress is effectively mutualized. This is not a novel risk\footnote{Similar dynamics already exist in single curator platforms such as Aave, and in TradFi when several funds hold the same bond and one becomes a forced seller} but modular vaults make these shared-liquidity channels more transparent and repackage them at the curator layer.

\begin{figure}[!htbp]
\centering
\includegraphics[width=\textwidth]{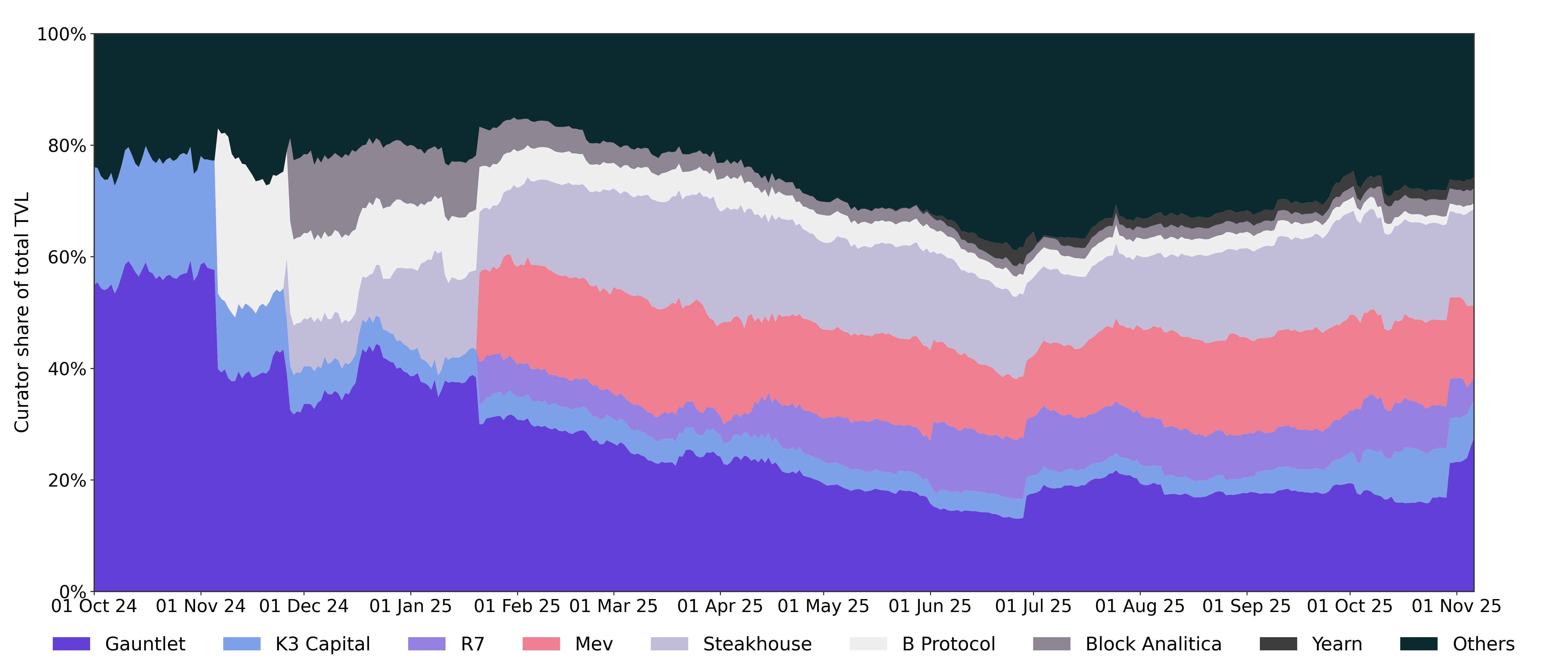}
\caption{Share of total curator-managed TVL among major curators.}
\label{fig:share_tvl_curators}
\end{figure}

Before examining overlap, tail dependence, and centrality, we document the scale of this layer. Figure \ref{fig:share_tvl_curators} shows that a handful of entities account for a large fraction of curated TVL. Across all analyzed risk curators, the total TVL amounts to approximately \$7.27 billion, distributed heterogeneously among eight key entities. Gauntlet dominates the ecosystem with about \$2 billion (27.6\% of total TVL), followed by Steakhouse (\$1.29 billion, 17.8\%), MEV Capital (\$915 million, 12.6\%), and K3\,Capital (\$478\,million, 6.6\%). The right tail is long, but the head is concentrated, which implies that shocks to a small group of managers can affect a disproportionate share of modular credit. 
Medium-sized curators such as R7, Block Analitica, Yearn, and B Protocol collectively accounts for 35.4\% of total exposure, while a fragmented long tail of smaller curators makes up the remaining $\approx$ 20\% of curated TVL.

\subsection{Portfolio structure and concentration}

\begin{figure}[!htbp]
\centering
\includegraphics[width=\textwidth]{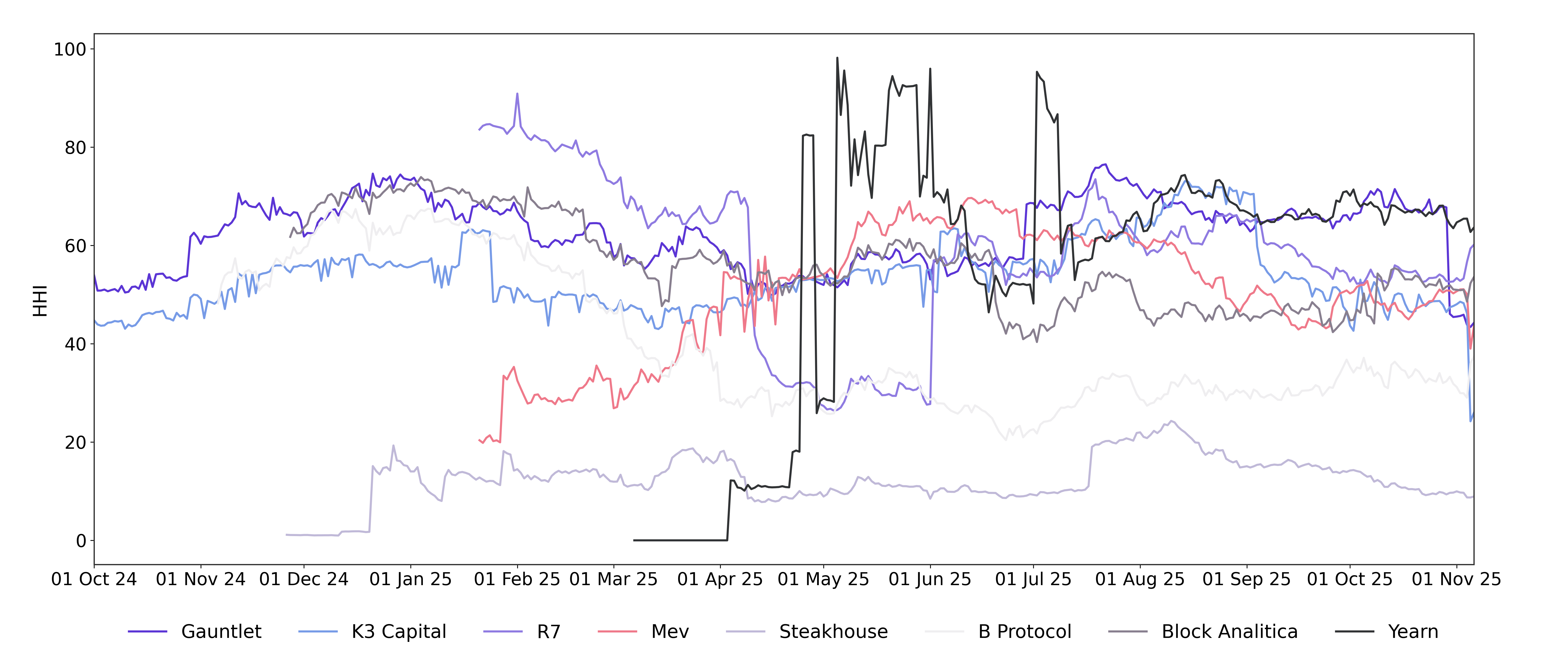}
\caption{Volatile-token exposure across curators.}
\label{fig:volatile_exposure_curators}
\end{figure}

\begin{figure}[!htbp]
\centering
\includegraphics[width=\textwidth]{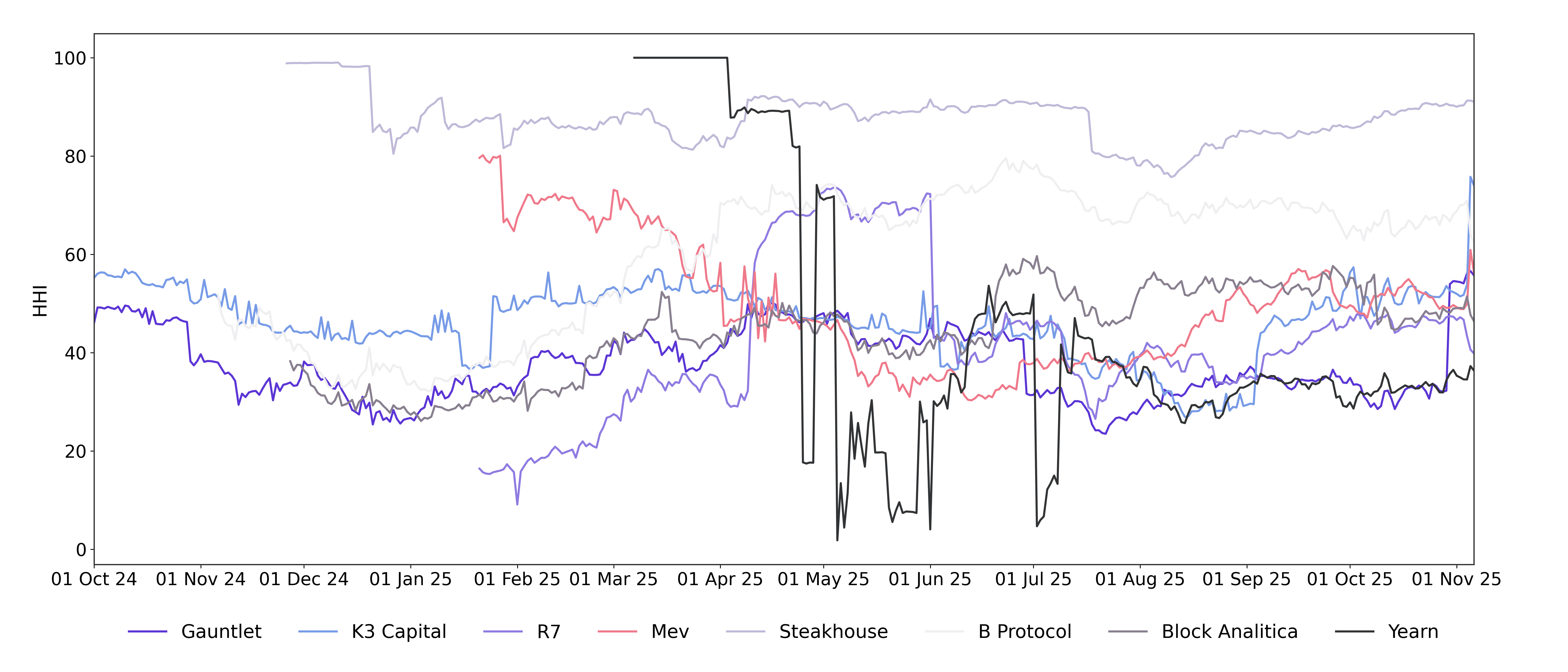}
\caption{Stablecoin exposure across curators.}
\label{fig:stablecoin_exposure_curators}
\end{figure}

\begin{figure}[!htbp]
\centering
\includegraphics[width=\textwidth]{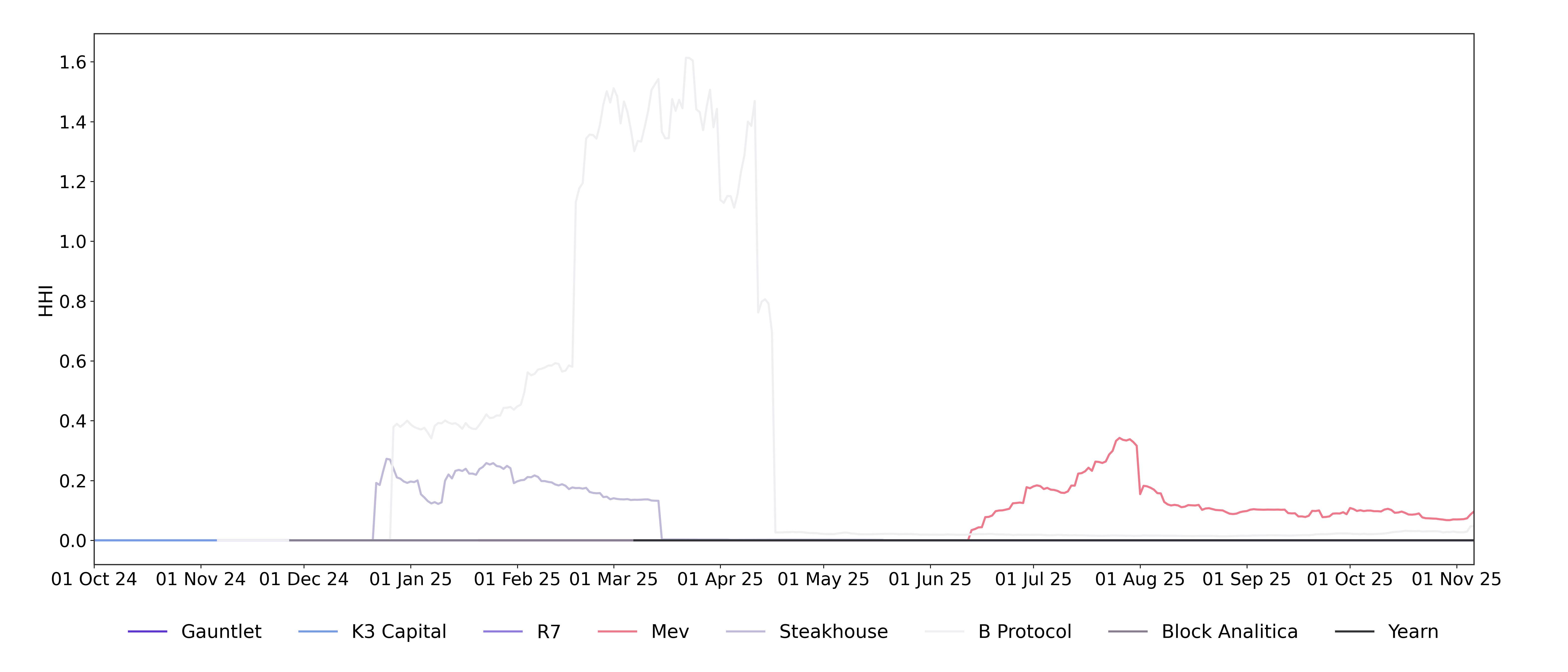}
\caption{Commodity (gold) exposure across curators.}
\label{fig:gold_exposure_curators}
\end{figure}

Figures~\ref{fig:volatile_exposure_curators}–\ref{fig:gold_exposure_curators} trace the evolution of curators’ portfolio composition and concentration.  
The volatile-token share in Figure~\ref{fig:volatile_exposure_curators} summarizes the fraction of each curator’s portfolio allocated to non-stable assets. For each curator, we compute this share as TVL deposited into volatile-asset markets divided by the aggregate TVL across all vaults they curate (including both stable and volatile positions). Over most of the sample, Gauntlet and R7 keep roughly 55  80\% of capital in volatile assets, effectively running high-beta credit whose performance is tightly linked to market conditions. Block Analitica starts in a similar range (around 60 – 70\%) but gradually de-risks towards 40 – 50\% by July 2025, while K3 Capital moves from about 50 – 60\% at launch to the low 20\% at the end of the sample. B Protocol reduced its volatile exposure from roughly 60\% to below 40\% after mid-2025. MEV Capital appears from early 2025 with 20 – 30\% volatile share, ramps up to around 60 – 70\% in mid-year, and then retraces towards 40\%. Yearn enters in early Q2 2025 with already high and jumpy volatile shares of 70 – 90\% and holds around 60\% from the mid 2025, reflecting its focus on ETH and restaking strategies. Steakhouse is the clear defensive outlier with almost stable volatile exposure below 20\%, reflecting its conservative collateral strategy.
The cross-section of volatile-asset shares, therefore mirrors, familiar segments in traditional credit markets with high-beta loan funds (Gauntlet, R7, Yearn, mid-2025 MEV Capital), intermediate-risk income funds (K3 Capital, Block Analitica), and low-beta cash-plus strategies (Steakhouse, B Protocol).

Stablecoin allocation in Figure~\ref{fig:stablecoin_exposure_curators} complements the volatile-share metric by tracking, for each curator and date, the fraction of aggregate curated TVL that is deployed into stablecoin markets.
Steakhouse exhibits the steadiest profile, spending most of the sample above 80\% and trending towards almost 90\% stablecoins by late 2025, which effectively makes it the system’s liquidity anchor. B Protocol similarly migrates from roughly 40 – 50\% to near‐full stablecoin allocation (around 80 – 95\%) after Q1 2025. R7 initially reduces its stablecoin share to the mid-20\%, from early 2025 progressively reallocates into stables, but in mid 2025 drops its exposure to around 40\% and keeps it until the end of the sample period. K3 Capital remains in a moderate band of roughly 40 – 60\% stables, with a sharp increase in the end of the sample to 80\%. Gauntlet moves in the similar direction, falling from about 50\% to the high-20\% by end-2024 and subsequently oscillating around 30 – 40\% (with a sharp increase at the end).
MEV appears in early 2025 with a predominantly stable allocation (around 70 – 80\%), then rotates into more volatile positions and settles near 40 – 50\% stables by year-end. Block Analitica maintains mixed exposures in the 30 – 60\% range, with a gradual drift towards higher stablecoin weights over time. Yearn enters in Q2 2025 with near-100\% stablecoin exposure, experiences a brief period of highly volatile composition as new vaults are launched and rebalanced, and then converges to a mixed profile of roughly 30 – 40\% stables.
Taken together with the volatile-share figure, this pattern traces a continuum of curator risk appetites: Gauntlet, R7, Block Analitica, MEV Capital, and Yearn operate as high- to medium-beta strategy managers, whereas Steakhouse and B Protocol function as low-volatility liquidity custodians, mirroring the segmentation between loan funds and cash-plus vehicles in traditional credit markets.

Finally, commodity exposure in Figure~\ref{fig:gold_exposure_curators} remains marginal across all curators, typically below 5\% of TVL and primarily held through synthetic or tokenized gold instruments. These allocations appear episodic and experimental, serving as hedging or diversification probes rather than strategic positions. Their low persistence suggests that, unlike in traditional multi-asset portfolios, commodities currently play no structural role in DeFi risk management. This absence highlights both the dominance of crypto-native beta and the underdeveloped linkage between DeFi credit and real-asset collateral.

\subsection{Systemic topology and tail dependence}

\begin{figure}[!htbp]
\centering
\includegraphics[width=\textwidth]{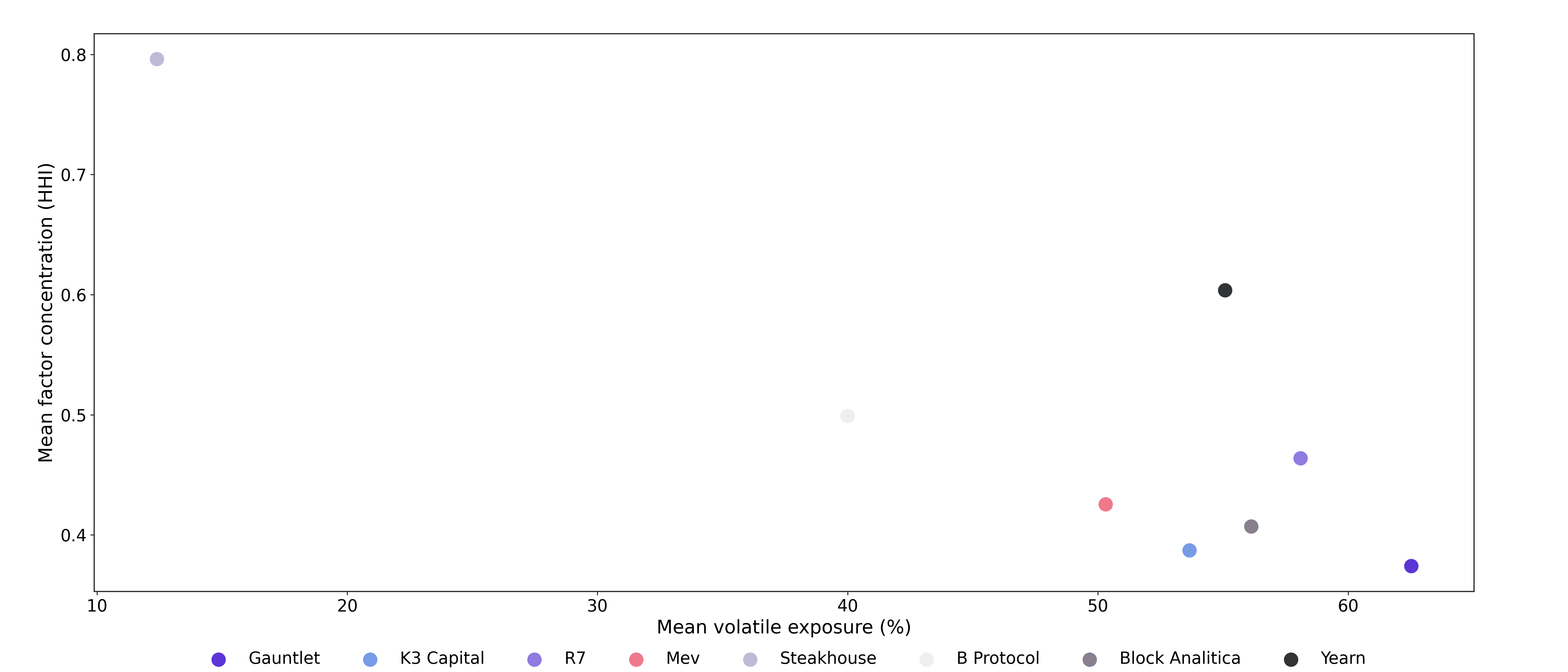}
\caption{Mean token concentration vs. volatile exposure.}
\label{fig:systemic_risk_map_curators}
\end{figure}

Figure~\ref{fig:systemic_risk_map_curators} summarizes the cross-sectional risk profile of curators by plotting the average volatile-asset share on the horizontal axis and the average factor-level HHI on the vertical axis. Thus, it jointly captures portfolio beta and diversification. The factor HHI is computed after aggregating tokens into underlying risk families, e.g., ETH and its liquid-staking derivatives, BTC and wrapped BTC, USD stablecoins, so that concentration reflects exposure to common risk drivers rather than to individual tickers. Rightward movement in the figure, therefore, indicates higher market beta, while upward movement indicates greater dependence on a small set of risk factors and, thus, lower capacity to absorb shocks.

Steakhouse occupies the upper-left region of the map, with a volatile share below 15\% and a factor HHI close to 0.8. This configuration corresponds to a low-beta balance sheet that is nonetheless highly concentrated in a single (USD-stable and RWA) factor. In traditional finance terms, this position is akin to a large money-market or government-bill fund. It is structurally long high-quality, low-volatility factor which offers liquidity and principal preservation rather than cross-factor risk-sharing. As such, Steakhouse functions as a structural stabilizer within the ecosystem.

B Protocol and Yearn lie further to the right and somewhat lower on the vertical axis. B Protocol exhibits an average volatile share of roughly 40\% and a factor HHI around 0.5, whereas Yearn combines a volatile share of about 55\% with a factor HHI near 0.6. Both curators load meaningfully on market risk while remaining concentrated in a limited set of factors, similar to concentrated sector or factor ETFs in regulated markets that take sizable directional bets on a few correlated risk premia. In stress scenarios, deleveraging in their vaults would transmit shocks strongly within those specific segments, much as redemptions from a sector fund amplify volatility within a narrow slice of the equity or credit segment. 

Gauntlet, K3 Capital, MEV Capital, R7, and Block Analitica form a cluster in the lower-right quadrant, with volatile shares in the 50--60\% range and factor HHIs between approximately 0.37 and 0.47. These curators operate with higher beta, but their exposures are distributed across several risk families, so they intermediate aggregate market risk without being dominated by a single factor. They are positioned to transmit broad market moves, but also to absorb idiosyncratic shocks in any one segment.

Taken together, the cross-section reveals a barbell structure in systemic roles: a highly factor-concentrated but low-beta liquidity provider (Steakhouse), a pair of factor-concentrated high-beta curators (Yearn and B Protocol), and a group of high-beta yet more factor-diversified managers. This mirrors patterns observed in emerging credit and bank lending markets prior to institutional standardization, where the system is dominated by a few quasi-money-market intermediaries at one end and a set of aggressive, lightly diversified risk-takers at the other, with relatively few genuinely balanced intermediaries in between.

\begin{figure}[!htbp]
\centering

\begin{minipage}{0.48\textwidth}
  \centering
  \includegraphics[width=\textwidth]{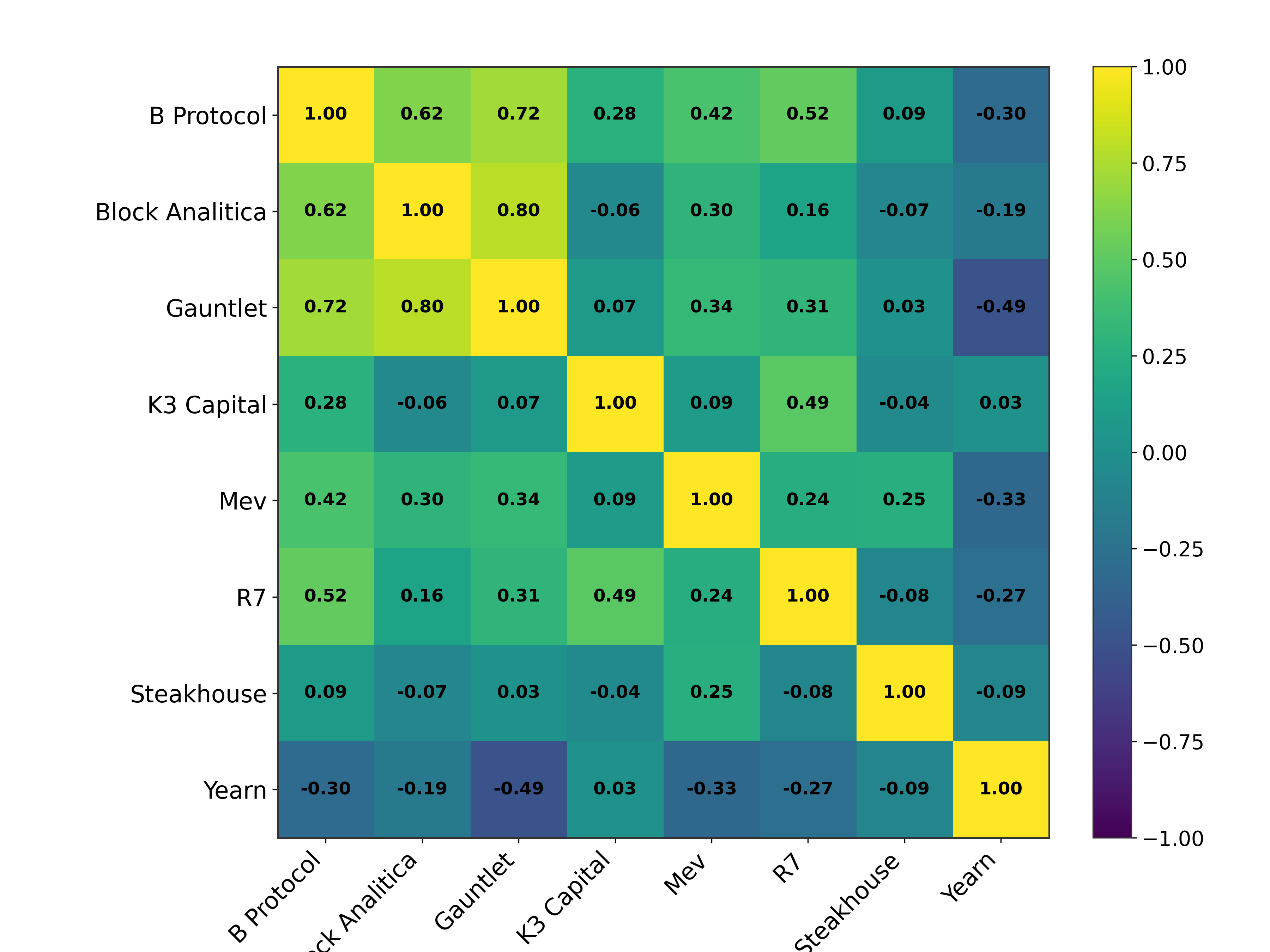}
  \vspace{0.3em}
  {\tiny TVL drawdown}
\end{minipage}
\hfill
\begin{minipage}{0.48\textwidth}
  \centering
  \includegraphics[width=\textwidth]{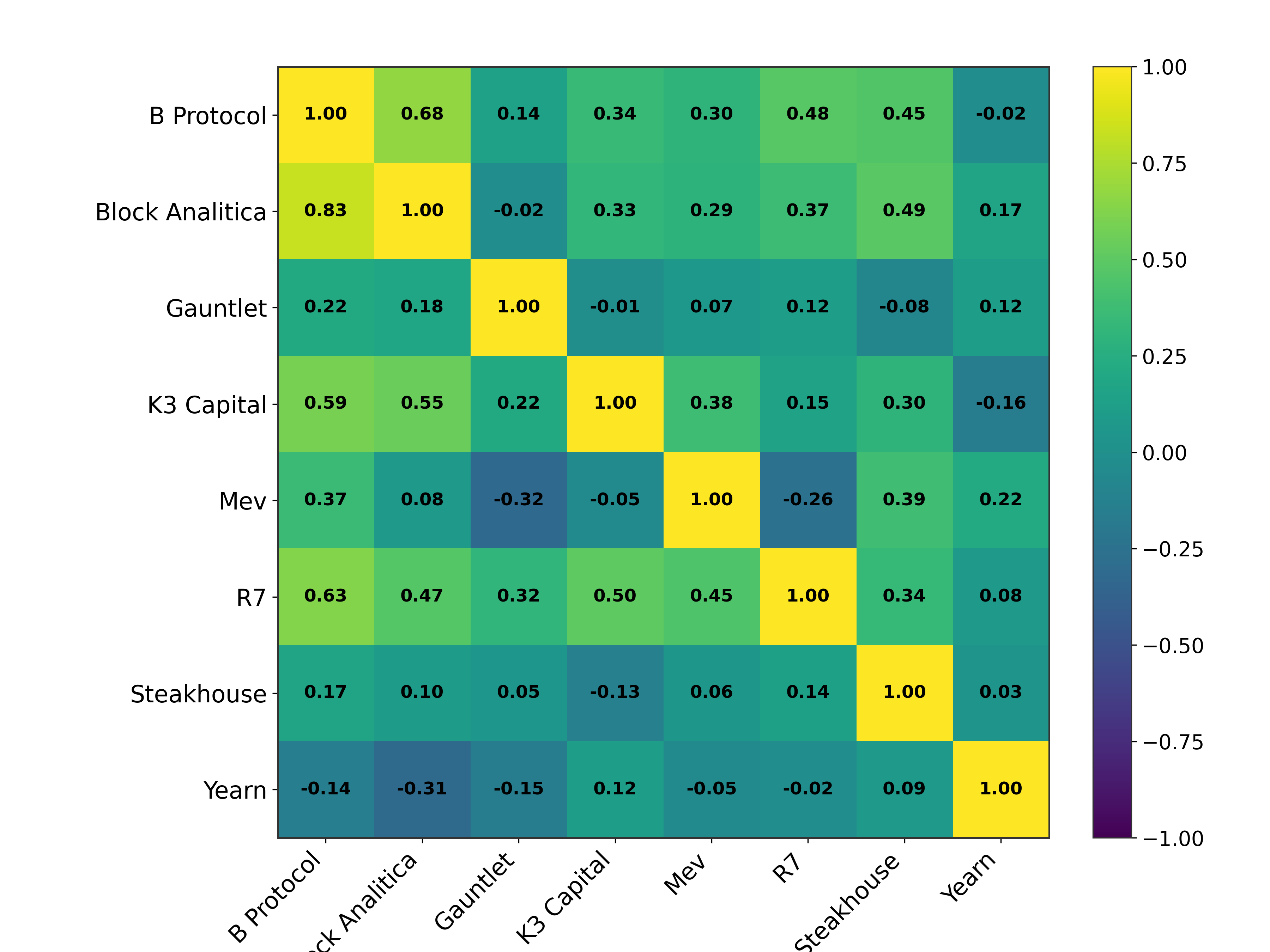}
  \vspace{0.3em}
  {\tiny Tail risk dependence}
\end{minipage}

\caption{Co-movement in liquidity stress across curators.}
\label{fig:comovement_curators}
\end{figure}

Figure~\ref{fig:comovement_curators} quantifies co-movement in liquidity stress across curators. 
For each curator, we compute the normalized drawdown, i.e., the fractional distance from the running peak. Pairwise drawdown correlation is then capturing synchronous deleveraging over the full sample. To isolate downside co-movement, we further construct a conditional lower-tail correlation, restricting attention to days on which a curator’s daily log-TVL return lies in its bottom decile. In this setting, a left-tail shock corresponds to such extreme negative return realizations (large outflows or sharp TVL contractions), so the conditional correlation measures the degree to which curators tend to experience joint stress precisely when one of them is hit by a severe adverse shock.
Both statistics are TVL-based, hence, they reflect jointly (i) price moves of held tokens and (ii) net flows. Because token baskets differ across curators, interpretation must be conditioned on exposure mix.

The drawdown matrix in the left panel of Figure~\ref{fig:comovement_curators} reveals a tightly connected core.
B Protocol, Block Analitica, and Gauntlet load strongly on the same liquidity cycle, with pairwise drawdown correlations of 0.72 (B Protocol–Gauntlet), 0.80 (Block Analitica–Gauntlet), and 0.62 (B Protocol–Block Analitica). These links are consistent with overlapping exposure to core DeFi collateral (ETH, LSTs, major stables) and broadly similar risk management responses. 
K3 Capital and MEV Capital exhibit intermediate correlations with the core, e.g., B Protocol–MEV at 0.42 and K3–R7 at 0.49, suggesting partial coupling but greater heterogeneity in mandate and execution. At the periphery, Yearn displays weak or negative correlations with most peers, e.g., -0.49 vis-à-vis Gauntlet and -0.30 vis-à-vis B Protocol, consistent with higher exposure to market beta and less driven by the stablecoin-liquidity cycle. 
Since its TVL responds primarily to market beta, stable-heavy curators may experience net inflows in risk-off periods, generating opposite-signed drawdown co-movement. Steakhouse shows near-zero drawdown correlations with the cluster, e.g., 0.09 with B Protocol and 0.25 with MEV Capital, behaving as a cash-management facility rather than an amplifying node.

The conditional tail-state dependence matrix in the right panel preserves the same hierarchy, but is more selective. 
Tail co-movement is highest for B Protocol–Block Analitica (0.68) and remains elevated for links between B Protocol and K3 Capital (0.59) and between B Protocol and R7 (0.63), indicating that when these curators experience left-tail TVL shocks, peers with similar collateral baskets tend to deleverage at the same time.
In contrast, MEV Capital displays small or negative tail correlations with Gauntlet and K3 Capital are small or even negative, e.g., -0.32 vs. Gauntlet, suggesting more orthogonal positioning or cross-chain diversification that dampens joint liquidations. Yearn again exhibits low tail dependence with the core cluster (mostly below 0.20 and negative for B Protocol and Block Analitica), consistent with its more volatile-token-intensive profile and distinct flow dynamics.

Overall, the heatmaps point to a modular but connected architecture. A small group of tightly coupled curators that can transmit liquidity stress efficiently, surrounded by stable-heavy or cross-chain allocators that act as buffers. Because TVL aggregates both price and flow components, these dependence patterns should be interpreted jointly with the composition measures in Figures~\ref{fig:volatile_exposure_curators}–\ref{fig:stablecoin_exposure_curators}. Low or negative correlations for Yearn relative to stable-dominant curators reflect opposite beta exposures rather than superior insulation per se, and suggest that shocks propagate along different channels. In practice, however, these channels are likely to track vault-level risk tiers rather than undifferentiated curator balance sheets, so a more granular decomposition by vault risk profile is a natural next step for further analysis.

Network analysis provides a structural view of the channels through which liquidity shocks can propagate across curators.  
To quantify interconnections, we model the system as a weighted undirected graph $G = (V,E)$, where each node $i \in V$ denotes a curator and each edge $(i,j) \in E$ summarizes the extent to which curators $i$ and $j$ hold the same underlying positions.
The edge weight $w_{ij}$ is defined as the share of TVL that is simultaneously invested in the same asset pools by both curators, divided by the smaller of the two total TVLs. Hence, $w_{ij}= $1 indicates that the smaller curator is fully nested inside the larger one, whereas $w_{ij}= $0 indicates no common exposures.  
To focus on economically relevant links, we retain only edges with $w_{ij} \geq$ 0.15, so that the resulting network reflects material portfolio overlap rather than incidental coincidences.
For each curator, we compute three standard centrality measures: 

\begin{enumerate}
\item Degree centrality counts the number of counterparties with which curator $i$ has significant overlap, normalized by the maximum possible number of links. It captures how many distinct channels a liquidity shock at $i$ can travel through.
\item Betweenness centrality measures how often curator $i$ lies on the shortest weighted paths between other pairs of curators, identifying nodes that act as bridges or conduits for shock transmission across otherwise weakly connected parts of the network. 
\item Eigenvector centrality assigns high scores to curators that are connected to other highly connected curators. It is a measure of systemic importance, in the sense that a curator is influential if it is embedded in the core of the overlap network.
\end{enumerate}

\begin{figure}[!htbp]
\centering
\includegraphics[width=\textwidth]{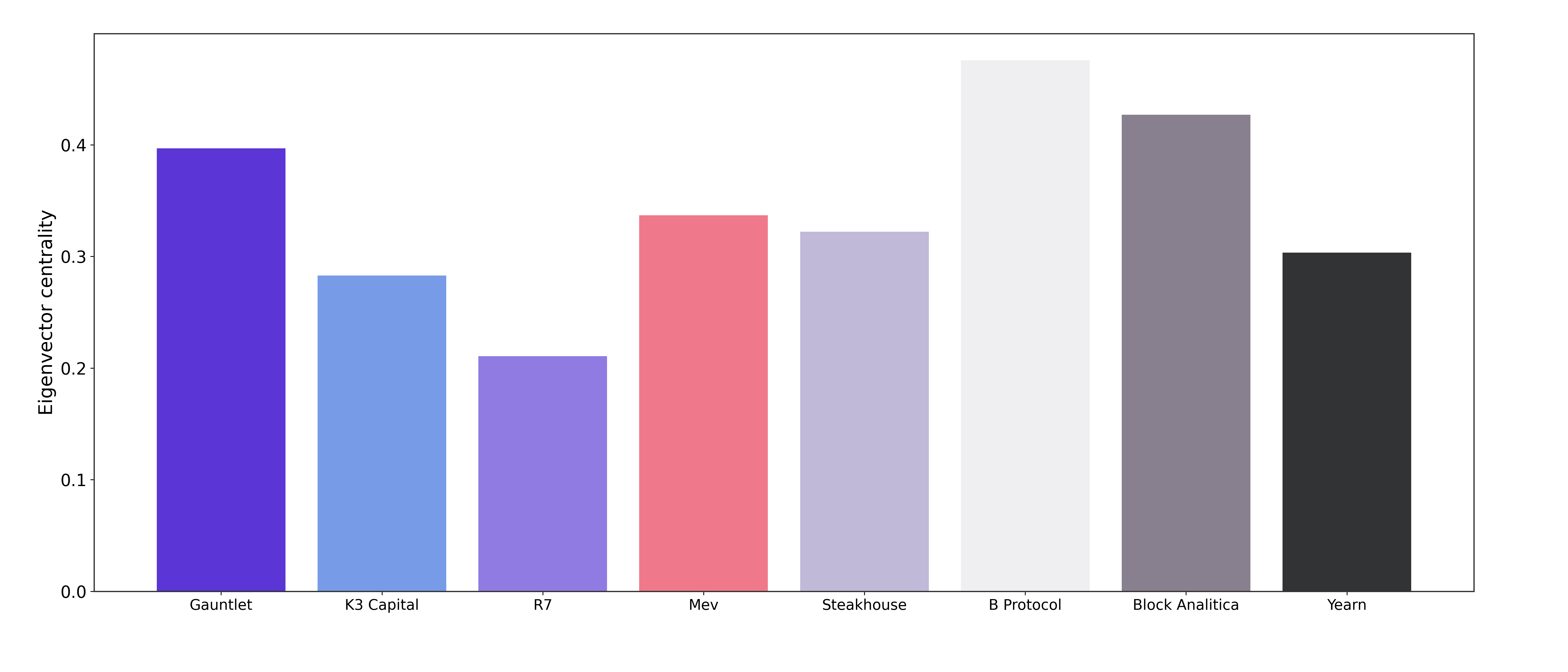}
\caption{Eigenvector centrality of curators in the liquidity-overlap network.}
\label{fig:systemic_importance_curators}
\end{figure}

Centrality metrics in Figure~\ref{fig:systemic_importance_curators} reinforce the hierarchical, but not purely size-driven, structure of the curator network. B Protocol and Block Analitica, though mid-sized in absolute TVL, exhibit the highest eigenvector centrality (roughly 0.46 and 0.43, respectively), identifying them as structurally critical hubs whose portfolio adjustments are most likely to transmit system-wide through overlapping exposures.  
A second tier consists of Gauntlet, MEV Capital, and Steakhouse, with eigenvector scores in the 0.32 – 0.40 range, indicating that they are strongly connected to the core but somewhat less system-defining. K3 Capital, Yearn, and R7 exhibit lower centrality (around 0.21 – 0.30), forming a more peripheral layer whose balance sheets are partly insulated from shocks originating in the network core. 

Taken together, the centrality pattern suggests that systemic disturbances would primarily propagate through the B Protocol–Block Analitica core and its immediate neighbors, while lower-centrality curators at the periphery provide a partial liquidity buffer that dampens contagion.

\begin{figure}[!htbp]
\centering
\includegraphics[width=\textwidth]{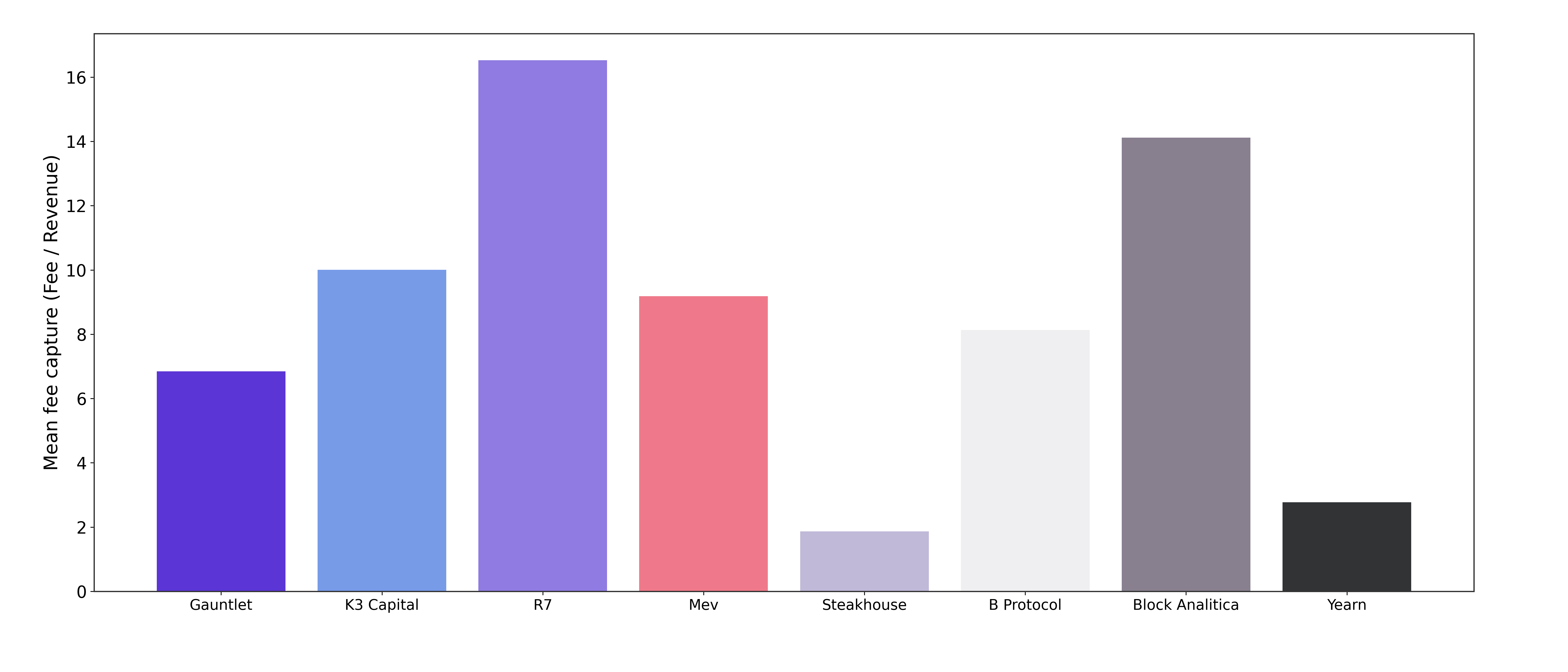}
\caption{Fee-to-revenue capture by curator.}
\label{fig:revenue_capture_curators}
\end{figure}

Finally, Figure~\ref{fig:revenue_capture_curators} reports the average share of accrued trading fees that is retained by each curator as net protocol revenue, i.e., the mean ratio of curator revenue to gross fees. This measure represents the curator’s effective profit margin on intermediation activity.
The results display striking cross-sectional dispersion. R7 and Block Analitica exhibit the highest fee capture, retaining roughly 16\% and 14\% of fees, respectively, with levels comparable to performance-fee spreads in actively managed hedge fund or structured-credit mandates.
Such margins are consistent with curators internalizing value from informational or algorithmic advantages, for example via proprietary execution logic, off-chain risk modeling, or privileged vault routing, but they do not uniquely identify this mechanism.  

K3 Capital and MEV Capital form an intermediate group, with average capture of about 10\% and 9\%, respectively, indicative of semi-automated optimization where value extraction is balanced against growth and rebate incentives. Gauntlet and B Protocol lie in the middle of the distribution (around 7\% and 8\%), consistent with delegated risk management providers that charge moderate spreads for model-based oversight.

At the lower end, Steakhouse and Yearn operate near cost recovery (below 3\%). Both follow architectures closer to public-utility risk management, i.e., maximizing protocol safety and liquidity depth. Their conservative, largely stablecoin-based mandates require less continuous rebalancing and active risk-taking, so a thinner compensation layer is consistent with a scale-driven rather than spread-driven revenue model. 

Two important caveats should be noted. First, headline take rates interact with heterogeneous risk profiles. Curators such as R7 and MEV Capital explicitly run more active, higher-beta strategies, so higher fee shares can be justified by higher gross yields. Many depositors focus on top-line net APY and do not fully adjust for risk, so a higher-yield, higher-risk vault can sustain a larger fee take while still appearing attractive relative to low-risk alternatives. Second, the protocol-level fee share we observe does not capture the economics of distribution and side agreements. In several cases, nominal curator fees are split with whitelabel distribution partners, e.g., lending protocols that host whitelabelled vaults, or upstream institutions, e.g., custodians and exchanges, and may be supplemented by incentive programs such as Morpho’s "Olympics". Without systematic data on such arrangements, Figure~\ref{fig:revenue_capture_curators} should, therefore, be read as a protocol-level upper bound on curator take rates, rather than a complete measure of curator-level profitability.

Overall, the dispersion in fee capture still points to an emerging differentiation of business models among DeFi curators. High-retention entities behave analogously to performance-fee funds in traditional asset management, monetizing active risk-taking and proprietary analytics, whereas low-retention entities resemble public-utility liquidity managers. A more precise mapping between take rates, risk intensity, and distribution economics is an important direction for future work.

\section{Transparency as a market discipline}
\label{sec:transparency}

Curator vaults replicate, in tokenized form, the leverage and liquidity engineering that has long been familiar from structured credit. They synthesize credit, duration, and basis exposures within modular smart-contract wrappers, allowing decentralized replication of (potentially) complex balance-sheet structures. In practice, however, effective duration and interest-rate sensitivity are only weakly disclosed and rarely priced explicitly by depositors, so duration risk is present but largely opaque for most market participants. A critical additional dimension is the vault’s permissible action set as enforced by immutable contract logic. Some designs, such as Morpho vaults, can allocate only to onchain, overcollateralized lending markets on the underlying protocol, whereas wrapper-style vaults may route deposits through EOAs into highly leveraged looping strategies or even off-chain assets. Understanding what a given vault can and cannot do with its collateral is, therefore, a first-order input into its risk assessment.
Table~\ref{tab:vault-strategies} summarizes representative strategy classes and their dominant risk channels.

\begin{table}[!htbp]
\centering
\caption{Representative vault strategies and associated risks}
\label{tab:vault-strategies}
\begin{tabular}{p{3.2cm}p{6.2cm}p{4cm}}
\toprule
\textbf{Strategy type} & \textbf{Mechanism} & \textbf{Primary risk} \\
\midrule
Stable/restaking loops 
& Recursive borrowing of stablecoins or LST assets to capture basis differentials or incentive spreads 
& Collateral correlation and liquidation cascades; aggressive oracle design shifting depeg risk onto lenders \\
Delta-neutral hedged 
& Synthetic dollar exposure via perpetual or option hedges 
& Hedge inefficiency and counterparty or oracle risk \\
Offchain credit/RWA 
& Tokenized private credit, treasuries, or invoice financing 
& Default risk and delayed loss recognition due to attestation lag and oracle latency \\
Aggregated yield routing 
& Meta-vaults reallocating capital across curators or protocols 
& Recursive exposure and dependency loops \\
\bottomrule
\end{tabular}
\end{table}

In the stable/restaking category, oracle design is a first-order parameter. Very aggressive or "hard-coded" oracle choices can effectively shift depeg and valuation risk from issuers onto lenders, even when nominal collateral appears stable. For offchain credit and RWA vaults, recent cases in which borrower defaults were only reflected in onchain NAV with a multi-week delay illustrate how attestation lags and oracle update frequency directly govern the timing of loss recognition.

These architectures elevate nominal yields by forming shadow collateral chains.
For instance, a recursive loop with an 80\% loan-to-value ratio achieves an effective leverage of $L = \frac{1}{1 - \mathrm{LTV}} = 5\times$, analogous to repo-style maturity transformation in traditional banking. In practice, sophisticated borrowers often operate at substantially higher effective leverage, especially where collateral is treated as near risk-free via hard-coded oracles, e.g., LSTs or yield-bearing stablecoins. Empirically, leveraged looping in assets such as weETH on Aave has frequently operated at health factors close to 1.02, corresponding to effective leverage well above 10$\times$.
\citet{lockedin2025risk} show that, for such leveraged borrowers, the tail of the loss distribution is highly sensitive to collateral volatility and liquidation penalty design, so that small changes in protocol parameters can materially shift the ruin probability.
The dynamic model of \citet{ChiuEtAl2024Fragility} formalizes an analogous mechanism for DeFi lending. When collateral valuations and rigid protocol-level haircuts interact, the system can tip between high- and low-lending equilibria even without explicit deposit contracts, so fragility arises from contract design more than from the use of volatile collateral itself. The Aave V2 CRV short-squeeze analyzed by \citet{HeimbachEtAl2024ShortSqueeze} shows that these vulnerabilities are not merely theoretical. Concentrating a large share of a token's market capitalization in a single protocol and relying on slow-moving DAO risk votes allowed a targeted short attack to generate substantial bad debt and emergency parameter changes. 
Such recursion increases return-on-equity in stable regimes but accelerates liquidation velocity under stress, which is precisely the dynamic that transformed pre-2008 collateralized lending into systemic contagion. The analytical role of curators is, therefore, critical as they render explicit the leverage and liquidity decisions that were previously embedded within monolithic pools.

Curators correspond functionally to the risk-bearing units of TradFi.  
Their organizational analogs clarify both strengths and vulnerabilities, as summarized in Table~\ref{tab:tradfi-vs-defi}. Recent initiatives by traditional rating agencies to assign public ratings to DeFi credit vehicles underscore this convergence and suggest that curator-managed vaults are beginning to be evaluated within familiar prudential frameworks.

\begin{table}[!htbp]
\centering
\caption{Comparative risk-governance structures in TradFi and DeFi}
\label{tab:tradfi-vs-defi}
\begin{tabular}{p{3cm}p{6cm}p{5.8cm}}
\toprule
\textbf{TradFi} & \textbf{Risk Management Logic} & \textbf{DeFi} \\
\midrule
CLO/ABS managers & Collateral selection under overcollateralization, IC/OC triggers, and waterfall rules & Vault curators configuring LTVs, liquidation thresholds, and eligible collateral sets, and absolute or relative caps on exposure to specific assets or markets\\
Money-market funds/UCITS & Diversification and liquidity ratio rules (e.g., 5/10/40 principle) & Stable-asset vaults (e.g., Steakhouse, B Protocol) with self-imposed composition and duration limits \\
Prime brokers/CCPs & Dynamic margining and collateral haircuts calibrated to volatility and liquidity & Gearbox "Ramping LT", Morpho adaptive parameter modules \\
Rating agencies & External credit assessment, stress-testing, and disclosure mandates & Credora scores; onchain risk-rating and attestation frameworks; traditional rating agencies beginning to rate DeFi credit structures \\
\bottomrule
\end{tabular}
\end{table}

At the protocol level, \citet{Oyeyemi2025DecentralizedCredit} propose a related evaluative framework for DeFi lending that combines collateral metrics, i.e., average LTV, collateral volatility, liquidation frequency, with a governance effectiveness index based on voter participation, token concentration, proposal execution, and crisis responsiveness. Using data for MakerDAO, Aave, Compound, Maple, and Goldfinch, they find that higher LTV ratios are associated with significantly more liquidation events, whereas stronger governance scores correlate with lower liquidation incidence and better handling of credit stress. Their results reinforce the idea that governance quality and risk disclosure are first-order determinants of credit outcomes in decentralized lending. In our setting, curator-managed vaults inherit protocol-level risk but also introduce their own governance and mandate heterogeneity, which calls for standardized, onchain disclosures at the curator and vault level.

A complementary perspective comes from the legal and consumer protection literature on credit scoring. \cite{PackinLevAretz2024BlackBox} argue that decentralized credit scores risk becoming "black box 3.0" systems since they are assembled from heterogeneous onchain and offchain data sources, executed through opaque algorithms, and embedded into smart contracts that can automatically restrict or price access to credit. In their analysis, the promise of decentralization does not by itself eliminate classic fairness and discrimination concerns. Instead, it shifts them into the interfaces between DeFi and TradFi, and to the governance of scoring models whose outputs may determine inclusion in or exclusion from lending markets. Insofar as curator-managed vaults rely on third-party credit scores for collateral eligibility, borrower whitelisting, or RWA underwriting, these “black box 3.0” issues become a first-order dimension of risk curation alongside solvency, liquidity, and market risk.

The distinction between TradFi and DeFi is institutional rather than conceptual. Where regulated finance enforces prudence ex ante via disclosure and capital requirements, modular DeFi enforces discipline ex post through transparency and open data. The equilibrium condition for sustainable modular credit is therefore clear: permissionless strategy creation must be counterbalanced by observable curation.
A credible framework for onchain transparency can emulate the informational rigor of regulatory reporting, without central supervision. At minimum, curator vaults should disclose the following onchain or via verifiable attestations:

\begin{enumerate}
    \item Asset-eligibility and issuer concentration: composition by token, sector, and per-issuer share of TVL.  
    \item Liquidity coverage ratio: an LCR-style metric that estimates the fraction of the vault that can be unwound under predefined stress scenarios, starting from onchain 1\% AMM slippage and extending to incorporate on- and offchain market depth, average daily volumes of collateral, and access to solver-mediated or just-in-time liquidity.
    \item Attestation cadence and signer quality: verification frequency and reputation scores for off-chain or RWA components.  
    \item Parameter reactivity: median time from a market shock to adjustment in LTVs, collateral caps, or interest parameters.  
    \item Rehypothecation map: dependencies across upstream lending markets and meta-vault interconnections.
    \item Fairness and access metrics: where credit scoring or whitelisting is used, summary statistics on score distributions, approval and rejection rates, binding limits across wallet cohorts or borrower types, to make potential disparate impacts observable onchain.
\end{enumerate}

Stress-test evidence on Compound indicates that, even within a single monolithic protocol, the combination of crypto collateral and leveraged stablecoin borrowing can generate meaningful default cascades under large price shocks~\citep{tovanich2023contagion}. In a modular, curator-driven architecture, these channels are multiplied since shocks can propagate not only across pools within a protocol, but also across vaults, chains, and curators sharing similar factor exposures. This amplifies the value of standardized disclosures on asset eligibility, liquidity coverage, and rehypothecation. At the same time, the "black box 3.0" concerns documented by \citet{PackinLevAretz2024BlackBox} imply that transparency must extend beyond balance-sheet quantities to the algorithms and criteria that gate access to credit. Making scoring dependencies and approval patterns observable where they interact with onchain vaults is, therefore, part of the same market-disciplining toolkit as publishing liquidity and concentration metrics.

Because ERC--4626 vaults already expose standardized accounting primitives, these transparency items can be embedded within token metadata or subgraph feeds. Aggregators, DAO treasuries, or independent rating agencies can then construct system-wide risk dashboards. In this configuration, transparency substitutes for regulatory supervision as it restores informational discipline without impeding innovation or curation speed. DeFi thereby converges toward a self-regulating equilibrium in which observable risk data anchors market trust.

\section{Conclusion}
\label{sec:conclusion}

The 2022 – 2025 period shows that DeFi did not merely add more lending protocols, it restructured credit into a two-layer system. At the base, lending infrastructures now range from fully pooled, governance-managed markets to fully modular, vault-instantiated markets. At the top, curator networks concentrate the economically meaningful risk choices and already display the same features that make dealer and CCP networks systemically relevant, i.e., portfolio overlap, tail co-movement, and a small group of highly central actors.

Conceptually, the system is shifting from closed, centrally planned underwriting at the protocol level to an open, permissionless underwriting regime at the curator layer, where the market collectively decides which assets and loans can be issued.
Because these curator networks are heterogeneous (they span stable-heavy, beta-heavy, and RWA strategies) but also interconnected, a purely laissez-faire posture would eventually reproduce the opacity and chain-like leverage that led to crises in traditional shadow banking.
The analysis of \citet{ChiuEtAl2024Fragility} shows that such fragility is ultimately a consequence of rigid contract rules interacting with expectations, rather than of decentralization per se, and the Aave V2 CRV short-squeeze case studied by \citet{HeimbachEtAl2024ShortSqueeze} illustrates how these design constraints can be exploited in practice once a protocol concentrates a large share of a token's market capitalization and relies on slow-moving governance updates.
The "black box 3.0" perspective of \cite{PackinLevAretz2024BlackBox} adds that opacity in credit scoring rails can similarly undermine the legitimacy of decentralized credit allocation if left unchecked.

The workable equilibrium for modular DeFi is, therefore, not more centralization, but more verifiable information about asset eligibility, liquidity coverage, attestation cadence, parameter reactivity, and rehypothecation maps, and, where relevant, the scoring-based filters that gate access to credit. Once these are public, users, DAOs, and even other curators can price vault risk on a comparable basis, bringing DeFi closer to the informational standards of money-market funds, UCITS vehicles, and rated structured credit, while preserving the permissionless, composable character that made vault-based DeFi attractive in the first place.

\bibliographystyle{apalike}
\bibliography{literature}

\end{document}